\documentclass[twocolumn,showpacs,prl]{revtex4}

\usepackage{graphicx}
\begin{document}

\title{Statistical Estimation of Quantum Tomography Protocols Quality}
\author{ $Yu.~I.~Bogdanov^{1}, G. Brida^{2}, M. Genovese^{2}, S.~P.~Kulik^{3}, E.~V.~Moreva^{4}, A.~P.~Shurupov^{2, 5}$}
\affiliation{1 Institute of Physics and Technology, Russian Academy of Sciences, Moscow, Russia}
\affiliation{2 INRIM, I-10135, Torino, Italy}
\affiliation{3 Faculty of Physics, Moscow State University, 119992, Moscow, Russia}
\affiliation{4 Moscow National Research Nuclear University "MEPHI", Russia}
\affiliation{5 Dipartimento di Fisica, Politecnico di Torino, I-10129, Torino, Italy}


\date{\today}

\begin{abstract}
A novel operational method for estimating the efficiency of quantum
state tomography protocols is suggested. It is based on a-priori
estimation of the quality of an arbitrary protocol by means of
universal asymptotic fidelity distribution and condition number,
which takes minimal value for better protocol. We prove the adequacy
of the method both with numerical modeling and through the
experimental realization of several practically important protocols
of quantum state tomography.
\end{abstract}
 \maketitle

\maketitle \textbf{Introduction.} Quantum states and processes being a fundamental tool for basic research, also become a resource for developing quantum technologies \cite{physrep}: this
demands for their characterization. At present
quantum state/process tomography serves as a standard instrument
\cite{tomo} for characterizing quality of preparation and
transformation of quantum states. Basically it includes a given set of unitary
transformations over the state to be reconstructed, then
the transformed state is measured and finally some computational
procedure applied to the measuring outcomes completes the state
reconstruction. The result is a state vector or a density matrix. In
order to compare different schemes the achieved result can
eventually be checked by calculating some distance measure, like
fidelity, that shows the quality of preparation/reconstruction of
the quantum state.
\\It is worth to mention that in real experiments the accuracy of the reconstruction depends on two types of uncertainties: statistical and instrumental ones.
If the total number of measurement outcomes (sample size or
statistics) is large enough, the instrumental uncertainties dominate
over fundamental statistical fluctuations caused by probabilistic
nature of quantum phenomena \cite{PRA_qutrit}. Practically the
required statistics, allowing to exclude statistical
 fluctuations, depends on the tomographic protocol itself and the total accumulating time needed for taking data. From this point of view it is of the utmost interest to point out simple and universal algorithms for the estimation of the chosen protocol on design stage before doing experiments as well as the sample size for desirable quality of the state reconstruction.
\\In this letter we propose a universal method for estimating the quality of any tomographic protocol based on discrete degrees of freedom and test it with well-known reconstruction protocols of polarization states of qubit pairs. Since it represents a paradigmatic example, here we restrict ourselves to polarization degrees of freedom only. Manipulations
with polarization qubits, qutrits and ququarts have been discussed widely in context of quantum state reconstruction and its optimization \cite{ kwiat, PRA_qutrit, PRA_ququart, D'Ariano, Rehacek, Kurtsiefer, Burgh,pas,nos}.

\textbf{Theory and numerical simulation} An arbitrary
$s$-dimensional quantum state is completely described by a state
vector in a $s$-dimensional Hilbert space when it is a pure state,
or by a density matrix $\rho$ for a mixed one. To measure the
quantum state one needs to perform a set of projective measurements.
According to Bohr's complementarity principle, it is impossible to
measure all projections simultaneously, operating with a single
representative of the quantum state only. So, first of all, one
needs to generate a set of copies of the state \cite{D'Ariano}. Then
for each measurement $j$ outcome rates $\lambda_{j}$ can be evaluated from amplitudes of a quantum process
\begin{equation}
\lambda_{j}=|M_{j}|^2.
\label{eq:prob}
\end{equation}
The amplitudes $M_{j}$ cannot be measured directly, but
they are linearly related to the state vector \cite{root-est}. For
an arbitrary protocol based on $m$ measurements the process
amplitudes can be represented as:
\begin{equation}
M_{j}=X_{j}c, \,\,\,\,\ j=1,2,..m \label{eq:ampl}
\end{equation}
$c$ being the state vector and $X_{j}$  a row of the so called
instrumental matrix $X$, which describes the entire set of
mutually-complementary measurements. The normalization condition for the protocol bounds the
total expected number  of events $n$ and the acquisition time $t_j$: $\sum _{j=1}^{m}tr(X_{j}^{+}X_{j}\rho)t_{j}=n$. We define
the row of the measurement matrix $B$ for a tomographic protocol as the direct
product $B_{j}=t_{j}X_{j}\bigotimes X_{j}^\star$, its size being
$m\times s^2$ and we assume $m \geq s^2$. By using the matrix $B$,
the protocol can be compactly written in the matrix form:
\begin{equation}
T = B\rho \label{eq:matrix}
\end{equation}
Here $\rho$ is the density matrix, given in the form of a column
(second column lies below the first, etc.). The vector $T$ of length
$m$ records the total number of registered outcomes. The algorithm
for solving equation (\ref{eq:matrix}) is based on the so called singular value decomposition (svd) \cite{svd}. Svd serves as
a base for solving inverse problem by means of pseudo-inverse (PI)
or Moore-Penrose inverse \cite{Penrose}. In summary, matrix $B$ can
be decomposed as:
\begin{equation}
B=USV^{+},\label{eq:svd}
\end{equation}
where $U$ ($m\times m$) and $V$ ($s^2\times s^2$) are unitary matrices and $S$ ($m\times s^2$) is a diagonal, non-negative matrix, whose diagonal elements are "singular values". Then (\ref{eq:matrix}) transforms to a simple diagonal form:
\begin{equation}
Sf=Q\label{eq:svd1}
\end{equation}
with a new variable $f$ unitary related to $\rho$ as $f=V^{+}\rho$
and a new column $Q$ unitary related to the vector $T$ as
$Q=U^{+}T$. We use this algorithm as a starting approximation for
maximal likelihood state reconstruction. By defining $q$ as the
number of non-zero singular values of $B$ we formulate two important
conditions of any tomography protocol, namely its completeness and
adequacy \cite{future}. The protocol is  supposed to be
informationally complete if the number of tomographically
complementary projection measurements is equal to the number of
parameters to be estimated; mathematically completeness means
$q=s^{2}$. Adequacy means that the statistical data directly
correspond to the physical density matrix (which has to be
normalized, Hermitian and positive). However, generally for mixed
state it can be tested only if the protocol consists of redundant
measurements, i.e. $m>q$.

\begin{figure}[h]
\includegraphics[width=0.95\columnwidth]{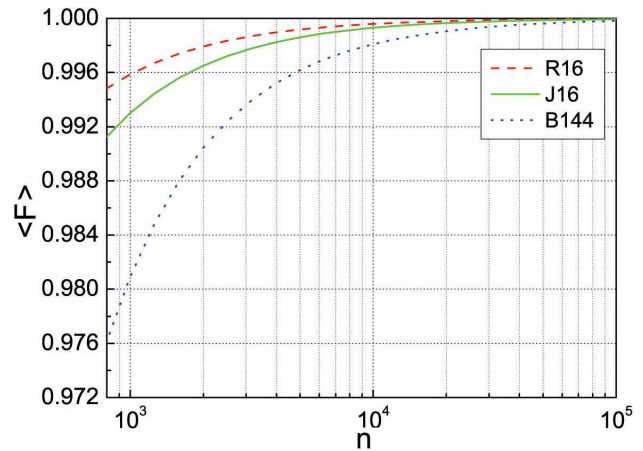}
\caption{(Color online) Dependence of the average fidelity on
number of registered events forming the sample.} \label{f:F_N}
\end{figure}

\begin{figure}[h]
\includegraphics[width=0.95\columnwidth]{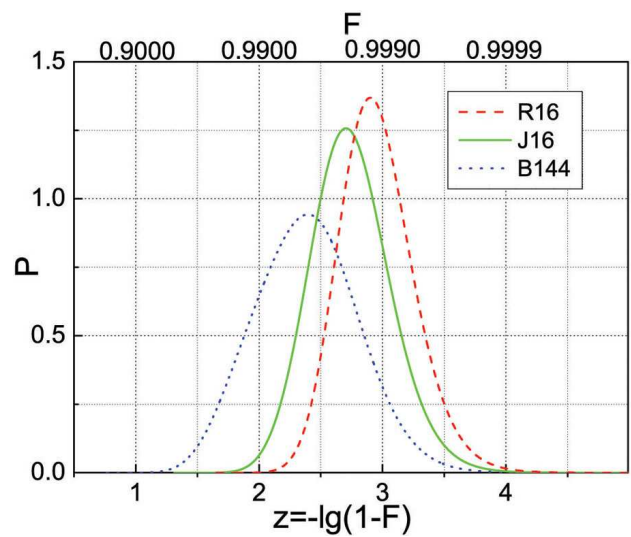}
\caption{(Color online) Density distribution  of the scaled fidelity
$z$  (lower abscissa) at $n=3\times 10^{3}$. Upper abscissa presents
regular fidelity.} \label{f:losses}
\end{figure}

PI provides with zero approximation for estimated parameters of quantum states. Optimization of these parameters can be done following the approach developed in \cite{PRA_qutrit} in the frame of maximal likelihood method. It is worth to stress that the reconstructed state vector extracted from the likelihood equation associates with finite statistics of the registered outcomes of an experiment and therefore takes random
values. The difference between the reconstructed state vector and
its exact value is caused by statistical fluctuations
\cite{instrumental}. The complete information matrix introduced in
\cite{PRA_qutrit, root-est} allows analyzing arbitrary functions on
fluctuating parameters of quantum states. One very relevant
function in the context of quantum state/process estimation is
fidelity $F = \left[ \textrm{Tr}\sqrt {\sqrt {\rho ^{( 0 )}} \rho
\sqrt {\rho ^{(0)}} } \right]^2$ where $\rho_{0}$ and $\rho$ being
the exact and the reconstructed density matrices correspondingly,
tending to identity for complete protocols with an unlimited
increase of a sample size.
\\Basically the problem of the state reconstruction splits into two scenarios. The first one relates to
the case when the reconstructed state is known in advance (like in
quantum cryptography). Then statistical comparison between exact
theoretical state and reconstructed one allows one either to reveal
the source of instrumental uncertainties or to be sure that such
uncertainties are small enough in comparison with statistical
fluctuations. The second case occurs when the reconstructed state is
unknown. Then using the approach developed below one should
calculate the statistical distribution of fidelity between unknown
exact state and reconstructed (by means of maximal likelihood
method) one. Fidelity will be either inside the uncertainty boundary
for small statistics or outside for large statistics (when
instrumental uncertainties prevail). In this case it is convenient
to analyze the so called fidelity loss $dF = 1-F$. When the
reconstruction accuracy is determined by a finite number of
representatives of the quantum state, fidelity loss is a random
variable whose asymptotic distribution can be represented as
\cite{Bogdanov09}:
\begin{equation}
1-F=\sum _{j=1}^{j_{\max } }d_{j} \xi_{j}^{2}
\label{eq:losses}
\end{equation}
 where $\xi _{j} \sim N\left(0,1\right)$ are independent and normally distributed random variables with zero mean values and unit
 dispersion, $d_{j}\sim\frac{1}{n} >0$, $j_{max}=2s-2$ for pure states, $j_{max}=s^2-1$ for mixed states. The coefficients $d_{j}$ can be extracted from
 the complete information matrix \cite{Bogdanov09}.
 Distribution (\ref{eq:losses})
 is a natural generalization of the $\chi ^{2}$ - distribution for which all $d_{j}=1$. As it follows from (\ref{eq:losses}) the average value of fidelity loss is:
\begin{equation}
\left\langle 1-F\right\rangle =\sum _{j=1}^{j_{\max } }d_{j}
\label{eq:average_losses}
\end{equation}
PI implies introduction of so called condition number $K$, which is
defined as the ratio between the minimal nonzero singular eigenvalue of $B$
and maximal one
\begin{equation}
K=\frac{b_{\max } }{b_{\min } }.
\label{eq:cond}
\end{equation}
$K$ determines the stability of the linear system (\ref{eq:matrix})
and therefore can be used as a practical quantifier for estimating
efficiency of the protocol: the lower $K$ the better the protocol.
If at least one of $s^{2}$ singular eigenvalues $b_{j}$ is close to
zero then the protocol
   becomes incomplete and $K\longrightarrow \infty$. The optimal value of $K$ would be unity that means uniform distribution of singular values \cite{future}.
\\The present approach allows to analyze arbitrary protocols of statistical reconstruction of quantum states.
As an example we have selected three popular protocols. The first,
suggested in \cite{kwiat} and dubbed J16 \cite{Burgh}, is suited for
reconstructing 4-dimensional polarization states, as photon pairs
degenerate in frequency generated in the process of spontaneous
parametric down-conversion (SPDC). In this protocol the projective
measurements upon some components of Stokes vector are performed for
each qubit in pair individually, so the 16 two-qubit measurements
are $HH$, $HV$, $VV$, $VH$, $RH$, $RV$, $DV$, $DH$, $DR$, $DD$,
$RD$, $HD$, $VD$, $VL$, $HL$, $RL$ where $H$, $V$, $R$, $L$, $D$
denotes horizontal, vertical, right and left circular and $45^o$
diagonal polarizations respectively. Here, for example, the
measurement setting $HR$ means measuring horizontal polarization on
the first qubit and right circular polarization on the second qubit
in pair. In our formalism the protocol can be represented by an
instrumental matrix with 16 rows $X_{j}$.
\\In the case of independent measurement of two qubits the projective measurements can be chosen arbitrarily. In \cite{Rehacek} measured qubits were projected on the states possessing tetrahedral
symmetry (R4). There are several works showing that due to
the high symmetry such protocol provides a better quality of
reconstruction \cite{Kurtsiefer, Burgh}. Let us stress that since
$m=s^{2}=16$ the adequacy of both R16 and J16 can not be
tested.
\\We also consider a protocol where the whole single-beam two-qubit state is subjected to linear transformations using two retardation plates
 \cite{PRA_ququart}(B144). The total number of
projective measurements is redundant and equals to 144, so the
corresponding instrumental matrix has 144 rows and protocol admits
adequacy testing, being $m>16$.
\\For these protocols we calculate condition numbers $K$ which take the following values: $K_{R16}=3$, $K_{J16}\approx 10$, $K_{B144}\approx 60$\cite{B144opt}. Thus, we expect that the symmetrical protocol R16 provides with better state reconstruction
quality \cite{Burgh}. We have checked this statement with numerical
simulations of each protocol applied to different two-qubit states
depending on the sample size.
\\As an example, let us consider the numerical reconstruction of the Bell
 state $\left| \Phi^{-}  \right\rangle=\frac{1}{\sqrt{2} } \left({\left| H_{1} H_{2}  \right\rangle} -{\left| V_{1} V_{2}  \right\rangle} \right)$.
 Fig. \ref{f:F_N} shows average fidelities, calculated according to (\ref{eq:average_losses}) as functions of sample size for each protocol.
 It turns out that the difference between curves disappears at sufficiently large sample size ($10^{5}$) \cite{instrumental},
 but for the same quality of state reconstruction the correct choice of protocol allows  using a smaller set of statistical data,
 i.e. finally reduces total acquisition time. Fig. \ref{f:F_N} shows that protocols are ranged in accuracy as following: R16, J16, and B144
 in complete agreement with the range given by condition number K.\
 Fig.\ref{f:losses} presents accuracy distributions calculated according to (\ref{eq:losses}) for the sample size $3\times 10^{3}$.
 For better visualization we transformed the abscissa scale into particular common logarithmic one as $z=-lg(1-F)$.
 Corresponding integers indicate the number of nines in fidelity recording.  The density distribution for the R16 protocol
 is narrower than the one for J16 and B144 and localizes in the region of lower losses or higher fidelities.
 The distribution for B144 is broader and lower in comparison with R16 and J16.
 Obviously, the narrower distribution of fidelity indicates a better reconstruction quality in the sense that
 the reconstruction procedure returns a better-defined state. Thus, Fig.\ref{f:losses}
 confirms our expectation based on estimation of condition number $K$: R16 achieves the best results.

\begin{figure}[h]
\includegraphics[width=0.95\columnwidth]{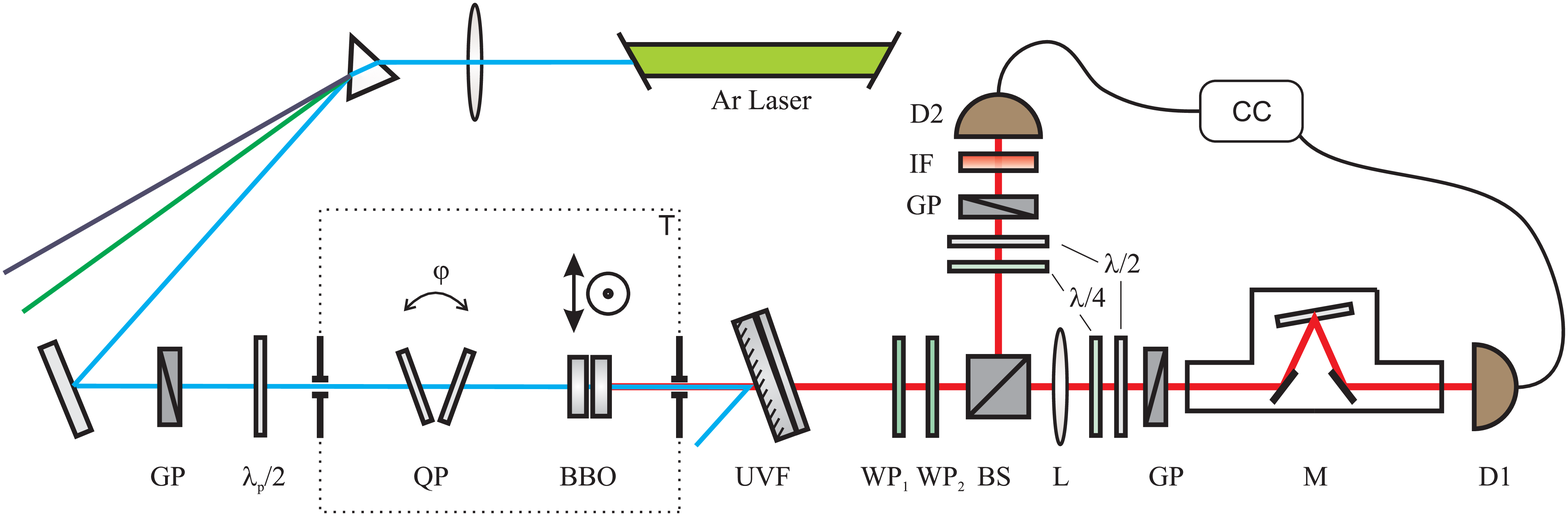}
\caption{(Color online) Experimental setup for different tomographic
reconstructions of  photon pairs with variable polarization
entanglement.} \label{f:setup}
\end{figure}

\textbf{Experiment.} We prepared a family of two-photon polarization
states which can be easily converted into both entangled (in
polarization) and factorized states
\begin{equation}
\left| \Psi  \right\rangle= \left(c_1{\left| H_{1} H_{2}  \right\rangle} +c_2e^{i\varphi }{\left| V_{1} V_{2}  \right\rangle} \right)
\label{eq:family}
\end{equation}
with real amplitudes $c_1$ and $c_2$. The set-up is schematically depicted in Fig.~\ref{f:setup}.
A cw argon laser beam at $\lambda=351nm$ pumps, after having
selected by a Glan-Thompson prism (GP) its horizontal polarization, two
type-I BBO crystals (1 mm) positioned with the planes that contain
optical axes orthogonal to each other. The halfwave plate placed in
front of crystals ($\lambda_p/2$) rotates the polarization of the
pump by the angle $\phi$ which controls real amplitudes $c_1$ and
$c_2$  in (\ref{eq:family}). The crystals are cut for collinear frequency non-degenerate
phase-matching around central wavelength 702nm. The relative phase
shift $\varphi$ in (\ref{eq:family}) is controlled by tilting quartz
plates QP. If $\phi=0$ we prepare the state
$\left| \Psi \right\rangle=\left| V_{1}V_{2} \right\rangle$, if
$\phi=22.5$ and $\varphi=3\pi/2$ then the state transforms to $\left| \Psi
\right\rangle=\frac{1}{\sqrt{2} } \left({\left| H_{1} H_{2}
\right\rangle} -{\left| V_{1} V_{2} \right\rangle} \right)\equiv \left| \Phi^{-}
\right\rangle$.

\begin{figure}[h]
\includegraphics[width=0.95\columnwidth]{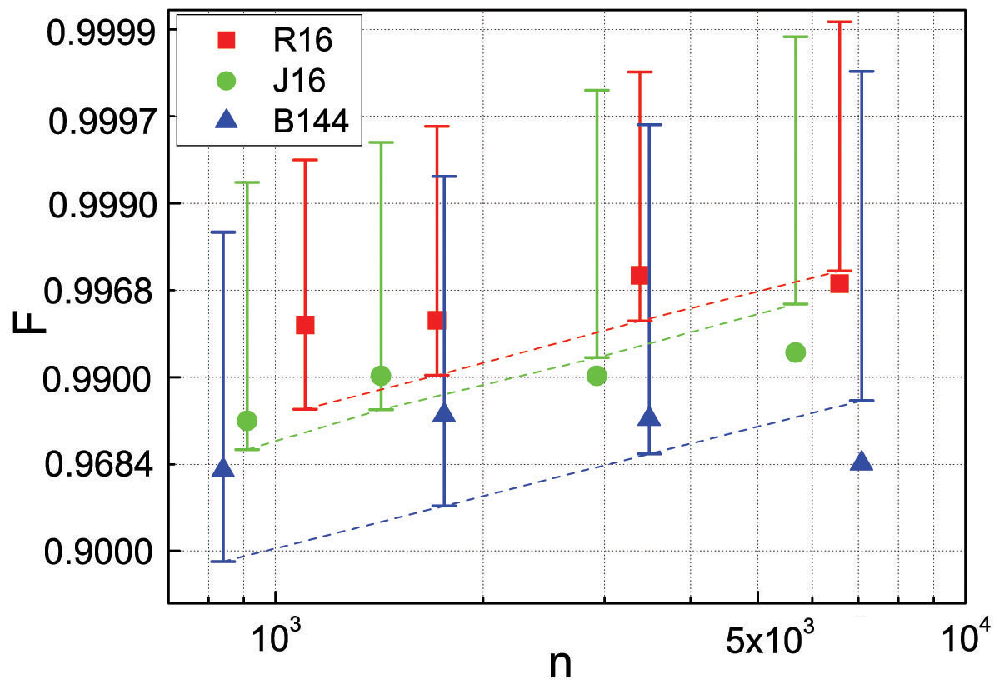}
\caption{(Color online) Reconstruction of $\left| \Phi^{-}  \right\rangle$-state. Vertical bars show 1\%- and 99\% - quantiles for fidelity distributions. Dotted lines connected lower bar ends indicate critical significant levels.}
\label{f:Psi}
\end{figure}

To maintain stable phase-matching conditions, BBO crystals and QP
are placed in a closed box heated at fixed temperature. The lens $L$
couples SPDC light into the monochromator M (with 1 nm resolution),
set to transmit "idler" photons at 710nm. The conjugate "signal"
wavelength 694 nm is selected automatically by means of coincidence
scheme. Zero-order wave plates ($\lambda/2,\lambda/4$) are used in
both J16 and R16  for arranging the projective measurement
set. In B144 these plates were removed and additional
achromatic quartz plates WP$_{1,2}$ (0.9183 mm,  0.9167 mm)
established the necessary measurement set at given wavelengths
\cite{PRA_ququart}.
\\For protocols described above the statistical reconstruction of prepared states (\ref{eq:family}) has been performed at given sample sizes.  As an example, Fig.(\ref{f:Psi}) shows calculated
widths of fidelity distributions at 1\%- and 99\%- quantiles for the
three protocols (see Fig.\ref{f:losses}) as well as the
experimentally reconstructed values for the specific state
$\left| \Phi^{-}  \right\rangle$.
The approach described above provides the ideal accuracy level for
quantum state reconstruction. It means that fluctuations of the
estimated quantum states certainly cannot lead to uncertainties
smaller than this limit. Presence of instrumental uncertainties
makes this level to be exceeded. Indeed, Fig.\ref{f:Psi} shows that,
beginning from some sample size, the experimental value of fidelity
falls out the theoretical uncertainty boundary shown as dotted lines
(corresponding to 1\% significance level which characterizes a given
protocol). This happens since instrumental uncertainties prevail
over the statistical ones and indicates that either state
preparation stage or measurement procedure were not performed
accurately enough. Incidentally, the absence of uncertainty bars for
data in Fig.4 derives from running once the corresponding protocol.
Repeated measurements would provoke appearance of a statistical
uncertainty, however respective theoretical uncertainty levels
should be re-scaled.
\\Definitely comparison between reconstructed states and fundamental
statistical fluctuations can serve to achieve precise adjustment of
the set-up, detection of the unapproved incursion into communication
channel, etc.

\textbf{Conclusion.} We have proposed and tested, both by numerical
calculation and experimental realization, a new estimation scenario
of tomographic protocols. Our results demonstrate the potentialities
of this method for a widespread application to experiments on
fundamental quantum optics and quantum technologies. Indeed, it can
provide in advance, based on condition number $K$, indications on
the uncertainty that can be reached for a certain set-up by a
tomographic scheme, providing  experimentalist with a tool for
choosing the best protocol based on available experimental resources (retardant plates, polarization
filters, etc.) and limited available time for data acquisition. Also
we would like  to stress that the developed method is quite general
and that it can be applied to any sort of quantum states and
measurement sets \cite{future}.

\textbf{Acknowledgments}. This work was supported in part by Russian
Foundation of Basic Research (projects 08-02-00741a, 10-02-00204a,
08-07-00481a), Leading Russian Scientific Schools (project
65179.2010.2), MIUR (PRIN 2007FYETBY) and NATO (CBP.NR.NRCL.983251).


\begin{thebibliography}{12}
\bibitem{instrumental} It holds if one can neglect instrumental uncertainties.
\bibitem{B144opt} Condition number for B144 can be optimized by chosing appropriate plates.
\end{thebibliography}
\end{document}